\newlength{\extralineskip}
\documentstyle[12pt]{article}

\begin{document}
\begin{titlepage}
\begin{flushright}
          \begin{minipage}[t]{12em}
          \large UAB--FT--391\\
                 May 1996
          \end{minipage}
\end{flushright}

\vspace{\fill}

\vspace{\fill}

\begin{center}
\baselineskip=2.5em

{\LARGE  GAMMA RAYS FROM SN1987A \\
DUE TO PSEUDOSCALAR CONVERSION}
\end{center}

\vspace{\fill}

\begin{center}
{\sc J.A. Grifols, E. Mass\'o, and R. Toldr\`a}\\

     Grup de F\'\i sica Te\`orica and Institut de F\'\i sica
     d'Altes Energies\\
     Universitat Aut\`onoma de Barcelona\\
     08193 Bellaterra, Barcelona, Spain
\end{center}

\vspace{\fill}

\begin{center}

\large ABSTRACT
\end{center}
\begin{center}
\begin{minipage}[t]{36em}
A light pseudoscalar coupled to two photons would be 
copiously emitted by the core of a supernova. Part
of this flux would be converted to $\gamma-$rays by
the galactic magnetic field. Measurements on the
SN1987A $\gamma-$ray flux by the Gamma-Ray
Spectrometer on the Solar Maximum Mission satellite 
already imply a bound on the
coupling $g < 3 \times 10^{-12}$ GeV$^{-1}$. The improved
generation of satellite-borne detectors, like EGRET 
or the project
GLAST, could be able to detect a pseudoscalar-to-photon
signal from a nearby supernova, for allowed values of $g$. 
\end{minipage}
\end{center}

\vspace{\fill}

Submitted for publication

\end{titlepage}

\clearpage

\addtolength{\baselineskip}{\extralineskip}

\section*{Introduction}

Pseudoscalar particles are often fundamental 
ingredients of Particle Physics models. Examples are
axions \cite{Weinberg-Wilczek} or Majorons
\cite{Chikashige-Gelmini}, coming from models with
spontaneous breaking of a Peccei-Quinn symmetry
\cite{Peccei} or of a global lepton symmetry, 
respectively. Other examples are
light bosons from extra-dimensional gauge 
theories \cite{Turok} and arions \cite{Anselm}. 
Pseudoscalar particles usually couple to two photons 
via the interaction lagrangian
\begin{equation}
  {\cal L} = \frac{1}{8} \ g \phi \ 
    \varepsilon_{\mu \nu \alpha \beta}
                           F^{\mu \nu} F^{\alpha \beta} .
\end{equation}

Limits on the coupling $g$ come  from laboratory experiments
and from cosmological and astrophysical observations. These
have been discussed and collected in \cite{Masso}. Further
astrophysical constraints have been recently examined 
by Mori \cite{Mori}.
The limits on $g$ depend on the mass $m$ of the pseudoscalar.
For very light masses, $m \leq 10^{-9}$ eV, the best 
constraints come from astrophysics (such constraints
are not interesting for axion models, where $g$ and $m$ are
related; but are of interest to other models where
$m \leq 10^{-9}$ eV does not imply an exceedingly small
coupling $g$). Mohanty and Nayak
\cite{Mohanty} considered the creation of a $\phi$
background due to strong magnetic fields in pulsars. Pulsar
signals propagating through this background show a time
lag between different modes of polarization. They found the 
limit
\begin{equation}
  g \leq 5.3 \times 10^{-11} \mbox{GeV}^{-1} ,
\end{equation}
valid for $m \leq 10^{-10}$ eV. Another way to constrain
$g$ for $m \leq 10^{-9}$ eV has been discussed by 
Carlson \cite{Carlson}.
He studied x-ray conversion of $\phi$ produced in
giant cores and found the following limit
\begin{equation}
  g \leq 2.5 \times 10^{-11} \; \mbox{GeV} ^{-1} ,
\end{equation}
using HEAO1 satellite data on $\alpha$-Ori x-ray emission.
A stringent bound on $g$ has been recently 
obtained by Krasnikov \cite{Krasnikov}: 
$\phi \rightarrow \gamma $ conversion
would lead to a large scale anisotropy of the x-ray
background. The observed isotropy leads to the estimation
\cite{Krasnikov}
\begin{equation}
  g \leq 1\times 10^{-11} \; \mbox{GeV} ^{-1}.
\end{equation} 
This author also considers similar effects to those 
studied by Carlson \cite{Carlson}. In addition he discusses 
polarimetric effects in the emission by magnetic white dwarfs,
and claims measurements may be sensitive to $g\sim 10^{-11}$
GeV$^{-11}$.

In the present paper, we show that it is possible to
improve these bounds on $g$ for $m \leq 10^{-9}$ eV. We
will use the fact that, should these pseudoscalars exist,
they are copiously produced when a supernova occurs. 
For small $g$, these particles will stream away 
from the supernova core,
without further interactions. In their path, the 
galactic magnetic field will convert a fraction of 
the flux back 
to gamma-ray photons. At the time the neutrino burst from
SN1987A was observed, the
Gamma-Ray Spectrometer (GRS) on the Solar Maximum Mission 
(SMM) satellite was operative and did not observe such a 
gamma-ray signal. We use this null result to set the
stringent limit
\begin{equation}
 g < 3 \times 10^{-12} \mbox{GeV}^{-1}.
\end{equation}
The next generation of detectors, like EGRET on the
Compton GRO satellite or the project GLAST, could
be able to detect a $\gamma -$ray signal due to $\phi$
conversion from a nearby supernova collapse, for
values of $g$ allowed by our analysis.

\section*{Pseudoscalar production}

Immediately after collapse, $\phi$ is produced in the
hot and superdense core of the supernova. In what 
follows, we take a
core temperature $T = 60$ MeV, proton number density
$n_p = 1.4 \times 10^{-38}$ cm$^{-3}$ and radius 
$R_c = 10$ km. As has been discussed in \cite{Turner},
a source of uncertainty arises because we do not know
which is the equation of state at supernuclear densities.
To estimate the uncertainty, we will follow that reference and 
change the central density by a factor of 2. All the
other parameters will change correspondingly. For example, the
core temperature changes by a factor 1.6, if one assumes
an isentropic collapse. Notice that this allows
the core temperature to span the conventional range of
$T\sim 30$ MeV $- \; 100$ MeV. Since $\phi$ couples to 
the electromagnetic field, the relevant interactions
will involve electrons and protons. The Fermi momentum
of electrons and protons is $p_F = 320$ MeV. For the
relativistic electrons, $p_F/T \gg 1$, thus they are 
extremely degenerate. As to the non-relativistic 
protons, $\frac{p^2_F}{2m_p} / \frac{3T}{2} \simeq 0.6$,
and hence they are only moderately degenerate.

Let us start considering the Primakoff process 
on protons, $p \gamma \longrightarrow p \phi$, that 
leads to $\phi$ creation. 
The number of pseudoscalars produced per 
unit volume and per unit time, at temperature $T$ and 
with energy between $E_{min}$ and $E_{max}$ is
\begin{equation} \label{N(T)}
  N(T) = \int_{E_{min}}^{E_{max}} \sigma (\omega) v \ n_p 
  \ dn_{\gamma} (T,\omega) .
\end{equation}  
We integrate between $E_{min}$ and $E_{max}$ since  
gamma-ray detectors are only sensitive to a fixed energy 
band. In expression (\ref{N(T)}) $n_p$ and $n_{\gamma}$ 
are the number densities of 
protons and photons respectively, $v$ their relative 
velocity and the Primakoff cross section as a function 
of the photon energy $\omega$ is (see \cite{Raffelt86})
\begin{equation} \label{cross section}
  \sigma(\omega) = \frac{1}{8} \alpha g^2 
  \left[\left(1+\frac{\kappa ^2}{4\omega ^2} \right)
  \ln \left(1+\frac{4\omega ^2}{\kappa ^2} \right)-1 \right]. 
\end{equation}
The screening wave number $\kappa$ appears because of
the collective behavior of the plasma, which cuts off
the range of the Coulomb potential for 
scales larger than $\sim \kappa ^{-1}$. It is given by
\begin{eqnarray}
  \kappa^2    & = & \kappa_D^2 + \kappa_{TF}^2     \\
  \kappa_D^2  & = & \frac{4\pi \alpha n_p}{T}  \label{kD}\\
  \kappa_{TF}^2 & = & \frac{4\alpha}{\pi}p_F E_F ,
\end{eqnarray}
with $p_F$ and $E_F$ the Fermi momentum and 
energy of the electrons. The proton 
contribution in our case is $\kappa_D = 50$ MeV, 
where we are considering the protons
not degenerate (see below). The contribution of the
degenerate electrons is given by $\kappa_{TF} = 33$ MeV,
smaller than the proton contribution. Protons can move
in the plasma (or better said, in momentum space) more 
freely than the degenerate electrons and so they can screen 
charges more easily than the electrons.

In order to obtain $\sigma(\omega)$ in eq. 
(\ref{cross section}) one needs to know the
thermal average $<\left| F(q) \right|^2 >$, being $F(q)$
the usual form factor associated to the charge distribution
of the plasma. To make the thermal average we need
$p_{ij}^Q(r)$, the probability per unit volume 
of finding the charge $Z_j$ at a distance 
$r$ of the charge $Z_i$ \cite{Raffelt86}
\begin{equation}
  p_{ij}^Q(r) = \frac{1}{V} \left( 1-\frac{Z_iZ_j \alpha}{T}
                 \frac{\exp(-\kappa r)}{r} \right)
\end{equation} 
for an arbitrarily large volume $V$.
In addition to this charge correlation one also has to consider
the statistical correlation between fermions. The probability
$p_{ij}^S(r)$ of finding a fermion $j$ at a distance $r$ of the
fermion $i$ is \cite{Pathria}
\begin{equation}
  p_{ij}^S(r) = \frac{1}{V} \left( 1-\exp (-mTr^2)
                                             \right).
\end{equation} 
One may estimate the relative importance of the two
different types of correlation by evaluating the quotient
\begin{equation}
  \epsilon \equiv \frac{V^{-1}-p_{ij}^S(r)}
                       {V^{-1}-p_{ij}^Q(r)} 
            \; \; \; \mbox{at} \; \; \; r=\kappa ^{-1} . 
\end{equation}
Using the physical parameters in the 
supernova core we obtain $\epsilon < 10^{-5}$ for the
gas of protons. Therefore we can neglect the near
degeneracy of protons and consider only the charge
correlation. In this case, one gets  
expression (\ref{kD}).

One might also consider other mechanisms of pseudoscalar
production such as the Primakoff process on electrons,
$e \gamma \longrightarrow e \phi$, or processes involving
only nucleons in the initial state, $pn \longrightarrow
pn \gamma \phi$, where a virtual photon attached either to
a proton or to a virtual charged pion splits into the final
$\gamma$ and $\phi$. These processes would add to the 
one previously considered and would make our limit on $g$ 
more stringent. However, the process $e \gamma 
\longrightarrow e \phi$ is suppressed compared to 
$p \gamma \longrightarrow p \phi$ due to the extreme
degeneracy of the electrons; production of a pseudoscalar
entails a change of momentum $\Delta p_e \sim T$ which
is not allowed for the bulk of electrons in the Fermi sea
(one has to bear in mind that forward peaks, in which 
$\Delta p_e \approx 0$, are suppressed by the finite range
of the Coulomb potential in a plasma).
Processes like $pn \longrightarrow pn \gamma \phi$ are also
unimportant since they can be visualized as $\gamma \phi$
production by the nearly static electric field created by 
protons. This sort of production is clearly diminished 
by energy conservation.  

Finally, we can write the expression for the pseudoscalar
flux $\Phi _{\phi} $ on the Earth, at a distance $D=55$
kpc from SN1987A, considering $p \gamma \longrightarrow 
p \phi$ as the only production mechanism in the
supernova core:
\begin{eqnarray}
  \Phi _{\phi}  =  3 \times 10^{4} \; \mbox{cm}^{-2} 
           \mbox{s}^{-1} 
           & & \left(\frac{g}{10^{-11} 
           \mbox{GeV}^{-1}}\right)^2 
           \left( \frac{55 \mbox{kpc}}{D} \right)^2   
           \left( \frac{R_c}{10 \mbox{km}} \right)^3
           \left( \frac{n_p}{1.4 \times 10^{38} \mbox{cm}^{-3}}
           \right) 
           \left( \frac{T}{60 \mbox{MeV}} \right)^3 
           \nonumber \\  & & \times f(\xi ^2,E_{min},E_{max}) ,
\end{eqnarray}
where 
\begin{equation}
  f(\xi ^2, E_{min}, E_{max})  = 
          \frac{1}{2\pi} \int_{x_{min}}^{x_{max}} dx
          \frac{1}{e^x -1} \left[(x^2+\xi^2)\ln 
          (1+x^2/\xi^2) - x^2 \right] , 
\end{equation}
and being 
\begin{eqnarray}
    \xi   & = & \kappa /2T,  \\
  x_{min} & = & E_{min}/T,   \\
  x_{max} & = & E_{max}/T.
\end{eqnarray}

One can check that the energy drain by these
pseudoscalars during the collapse of the
supernova core is at least a factor 
$10^{-4}$ smaller than
the energy released in neutrinos, therefore we can
ignore the back reaction caused by $\phi$ emission 
on the supernova evolution.

\section*{Pseudoscalar conversion}

The mixing between the photon and low mass particles in 
magnetic fields leads to very interesting phenomena. Since
the pioneering work in \cite{Sikivie}, a variety of 
implications for laboratory experiments and
astrophysical observations have been
investigated (see references in \cite{Masso}).

For our purposes, we use the formalism developed in
\cite{Stodolsky}. The $\phi \rightarrow \gamma$
transition probability, for a beam traversing a transverse 
magnetic field $B_T$ after a distance $L$, is given by
\begin{equation} \label{probability}
  P\left( \phi \rightarrow \gamma \right) = \frac{1}{4}
                 g^2 \ B^2_T \ L^2 \ 
                 \left( \frac{\sin x}{x} \right)^2 ,
\end{equation}
with
\begin{equation}
  x  =  \frac{L}{2} \sqrt{\Delta^2_{osc} + g^2 B^2_T} 
\end{equation}
and
\begin{equation}
  \Delta_{osc}  =  \frac{\left| \omega ^2_p - m^2 \right|}
                        {2 \omega} ,
\end{equation}
where $\omega$ is the energy and 
\begin{equation}
  \omega ^2_p = \frac{4\pi \alpha n_e}{m_e} = 
     \left( 6.4 \times 10^{-12} \; \mbox{eV} \right) ^2
     \left(\frac{n_e}{0.03 \; \mbox{cm}^{-3}} \right)
\end{equation}
is the plasmon mass. We have normalized it to the mean 
electron density in the interstellar medium, $n_e \simeq
0.03$ cm$^3$.

A coherent effect, which implies 
$\sin x /x \rightarrow 1$, is obtained provided
\begin{equation}
  m \leq 10^{-9} \; \mbox{eV},
\end{equation}  
so that our constraint on $g$ will be valid only for such
small masses. In order to evaluate the probability in
(\ref{probability}), we need to specify the magnetic field
structure. We will adopt the model used in \cite{Carlson}, 
consisting of 
a toroidal $2\mu$G magnetic field. The coherence length
is about the order of several kiloparsecs \cite{Zeldovich}.
To be conservative we take a coherence
length $L = 1$ kpc. In the direction of SN1987A one has
\begin{equation}
  B_T = 2 \mu \mbox{G} \left(1-\sin ^2 l \cos ^2 b \right)^{1/2}
      \simeq 1 \; \mu \mbox{G} ,
\end{equation}
where the corresponding galactic coordinates have been used.

Summing up, we have
\begin{equation}
 P\left( \phi \rightarrow \gamma \right) = 3 \times 10^{-3} 
           \left( \frac{g}{10^{-11} \; \mbox{GeV} ^{-1}} \right) ^2 
           \left( \frac{B_T}{1 \; \mu \mbox{G}} \right) ^2
           \left( \frac{L}{1 \; \mbox{kpc}} \right) ^2
\end{equation}

\section*{Results and Conclusions}

At the time the neutrino burst from  SN1987A
reached the Earth, the satellite-borne GRS was on
duty measuring the incident $\gamma-$ray flux. This
measurement provides an observational limit on the
$\gamma-$ray flux coming from the supernova, and
consequently on the photons from supernova $\phi$
emission.
Indeed, data from SMM \cite{Chupp} in the energy 
band $E_{min} = 25$ MeV $< E < E_{max} = 100$ MeV 
imply that
\begin{equation}
  \Phi_{\phi} \ P(\phi \rightarrow \gamma) \ \Delta t
  < 0.6 \; \mbox{cm}^{-2}.
\end{equation}

The characteristic time over which the proto-neutron
star emits most of its gravitational energy and 
therefore the bulk of the hypothetical $\phi$ particles
is the diffusion time of neutrinos. Detailed stellar
evolution calculations \cite{Burrows} render 
about 10 seconds for this diffusion
time scale. To be conservative we shall take 
$\Delta t = 5$ s which is roughly the characteristic
signal decay time for the neutrino burst of SN1987A.
We thus find the limit 
\begin{equation} \label{limit}
  g < 3 \times 10^{-12} \; \mbox{GeV}^{-1}.
\end{equation}
Changing the physical parameters of the supernova
as previously discussed, we estimate an uncertainty
factor of 2 in the limit (\ref{limit}).

Finally, we would like to make some comments on the
energies of the photons from pseudoscalar conversion.
The expected spectrum has a mean energy value 
slightly bigger than $2.7 T \sim 160$ MeV (the energy
spectrum is not exactly that of a black body since
the Primakoff cross section increases with the energy).
Thus, it extends above the GRS cut-off at $E_{max} =
100$ MeV and, consequently, part of it would not 
have been  detected.
Better prospects could be expected with EGRET on 
the Compton GRO satellite, launched
in 1991, since it is able to detect $\gamma-$rays from
$E_{min} = 20$ MeV up to $E_{max} = 30$ GeV and therefore
is sensitive to the whole spectrum. In addition, the
EGRET detector is more sensitive than the GRS detector. A
pseudoscalar with $g$ allowed by equation (\ref{limit}) 
would possibly give a $\gamma -$ray signal in EGRET 
from a nearby supernova.
Even more promising is the project GLAST with detectors
to measure $\gamma -$rays in the energy range
from $E_{min} = 50$ MeV to $E_{max} = 100$ GeV, and a
factor 100 better in sensitivity than EGRET.

\section*{Acknowledgments}

We thank the Theoretical Astroparticle Network for support under
the EEC Contract No. CHRX-CT93-0120 ( Direction 
Generale 12 COMA ). This work has been partially supported by 
the CICYT Research Project Nos. AEN95-0815 and AEN95-0882. 
R.T. acknowledges a FPI Grant from the Ministerio de 
Educaci\'{o}n y Ciencia (Spain).

\end{document}